\documentclass[prx,twocolumn,superscriptaddress,citeautoscript,showpacs,amsart,longbibliography]{revtex4-2}

\usepackage{blindtext}
\usepackage{graphicx}
\usepackage{bm}
\usepackage{amssymb}
\usepackage{amsfonts}
\usepackage{braket}
\usepackage{color}
\usepackage{orcidlink}
\usepackage{hyperref}
\usepackage{float}
\restylefloat{table}
\usepackage{bibentry}
\usepackage{color}
\usepackage{multirow}
\usepackage{amsfonts}
\usepackage{amsthm}
\usepackage{graphicx}
\usepackage{multirow}
\usepackage{color}
\usepackage{bbold}
\usepackage{bm}
\usepackage{times}
\usepackage{amsmath,bm,amsfonts}
\usepackage{dcolumn}
\usepackage{graphicx}
\usepackage{latexsym}
\usepackage{ulem}
\newcommand{\bpm}{\begin{pmatrix}}
\newcommand{\epm}{\end{pmatrix}}
\newcommand{\ba}{\begin{eqnarray}}
\newcommand{\ea}{\end{eqnarray}}
\newcommand{\bd}{\begin{displaymath}}

\graphicspath{{figures/}}

\begin{document}
\title{Strain-tuned orbital-dependent electronic correlations in FeTe thin films}

\author{Hyunjee Song}
\affiliation{Department of Physics and Astronomy, Seoul National University, Seoul 08826, Korea}

\author{Sangjae Lee}
\affiliation{Department of Physics and Astronomy, Seoul National University, Seoul 08826, Korea}

\author{Keun-Yeol Park}
\affiliation{Department of Physics and Astronomy, Seoul National University, Seoul 08826, Korea}

\author{Jaehyun Park}
\affiliation{Department of Physics and Astronomy, Seoul National University, Seoul 08826, Korea}

\author{Suyoung Lee}
\affiliation{Department of Physics and Astronomy, Seoul National University, Seoul 08826, Korea}

\author{Yeonjae Lee}
\affiliation{Department of Physics and Astronomy, Seoul National University, Seoul 08826, Korea}

\author{Jinyoung Kim}
\affiliation{Department of Physics and Astronomy, Seoul National University, Seoul 08826, Korea}

\author{Jaeung Lee}
\affiliation{Department of Physics and Astronomy, Seoul National University, Seoul 08826, Korea}

\author{Celesta S. Chang\,\orcidlink{0000-0001-7623-950X}}
\affiliation{Department of Physics and Astronomy, Seoul National University, Seoul 08826, Korea}

\author{Younsik Kim\,\orcidlink{0009-0005-7928-9322}}
\email[Electronic address:$~~$]{leblang@snu.ac.kr}
\affiliation{Department of Physics and Astronomy, Seoul National University, Seoul 08826, Korea}

\author{Changyoung Kim\,\orcidlink{0000-0003-0555-2341}}
\email[Electronic address:$~~$]{changyoung@snu.ac.kr}
\affiliation{Department of Physics and Astronomy, Seoul National University, Seoul 08826, Korea}

\date{\today}

\begin{abstract}
Iron chalcogenides exhibit rich phenomena which are governed by orbital-dependent electronic interactions and strong electronic correlation. In particular, many studies have explored orbital selectivity in FeTe through Se doping. Here, applying tensile strain to thin films allows us to precisely control the system without other impurities that may arise from chemical doping to investigate the emergent behaviors in FeTe. Using angle-resolved photoemission spectroscopy, we observe a spectral weight transfer between $d_{\rm xy}$ and $d_{\rm z^{2}}$ orbitals, evidence of an orbital-selective Mott phase (OSMP). Beyond OSMP, we reveal hitherto unobserved strain-induced effects, distinct from chemical doping. The evolution of $d_{\rm xz}$ orbital demonstrates how electron hopping mechanism plays an important role in defining the electronic properties of the system. Our findings highlight a direct correlation between epitaxial strain and the evolution of electronic structures in FeTe.

\end{abstract}
\maketitle

\section{Introduction}

Strongly correlated systems with multiple interacting parameters can host a variety of novel quantum phases. However, the strong entanglement among these parameters also presents challenges in identifying the underlying microscopic mechanisms. A representative example is the iron chalcogenide (FeCh) family, which exhibits a rich phase diagram encompassing unconventional superconductivity, nontrivial topology, magnetism, nematicity, and an orbital-selective Mott phase (OSMP)~\cite{paglione2010high,kim2024orbital,zaliznyak2011unconventional,wang2017superconductivity,subedi2008density,greger2013emergence,zhang2018observation,zhang2019multiple,ishida2020novel}. One effective approach to explore these phases has been isovalent chalcogen substitution~\cite{yu2013orbital,peng2019observation,liu2010pi}. Yet, the decisive control parameter remains elusive, as the elemental substitution simultaneously alters both the Fe–Ch–Fe bond angle and the Fe–Ch bond length, each exerting distinct influences on different orbitals~\cite{yi2017role}. This intertwining accompanied by chemical substitution poses challenges to disentangle the effects of individual structural parameters and study the emergent properties in FeChs.

While quantum materials have traditionally been studied in bulk single-crystal form~\cite{yin2011kinetic,de2014selective,si2016high,nakajima2014normal,niu2016unifying,maheshwari2018heat,PhysRevB.88.115130}, epitaxial thin films offer an alternative route to separate these intertwined effects. In iron chalcogenides, for example, the Fe–Ch–Fe bond angle can be selectively tuned by epitaxial strain while keeping the bond length nearly constant, providing a unique tuning knob inaccessible in bulk substitution~\cite{seo2017origin}. Moreover, the strain can be continuously adjusted by changing the film thickness, as relaxation gradually occurs in thicker films. This tunability enables systematic control of orbital-dependent correlation effects that remain obscure in bulk systems.

In this work, we investigate the evolution of the electronic structure of FeTe thin films under tensile strain using {\it in-situ} angle-resolved photoemission spectroscopy (ARPES). We observe clear evidence of the collapse of the OSMP in the $d_{\rm xy}$ orbital of tensile-strained FeTe, accompanied by enhanced decoherence in the $d_{\rm xz}$ orbitals. This opposite trend for the $d_{\rm xz}$ orbitals sharply contrasts with that of bulk Fe(Te,Se), where Se substitution enhances the coherence in all $t_{\rm 2g}$ states~\cite{ieki2014evolution,liu2015experimental,huang2022correlation}. By leveraging the precise structural control available through epitaxial strain, our results establish a pathway to manipulate orbital-dependent electronic correlations in thin films.

\begin{figure}
\includegraphics[width=0.48\textwidth]{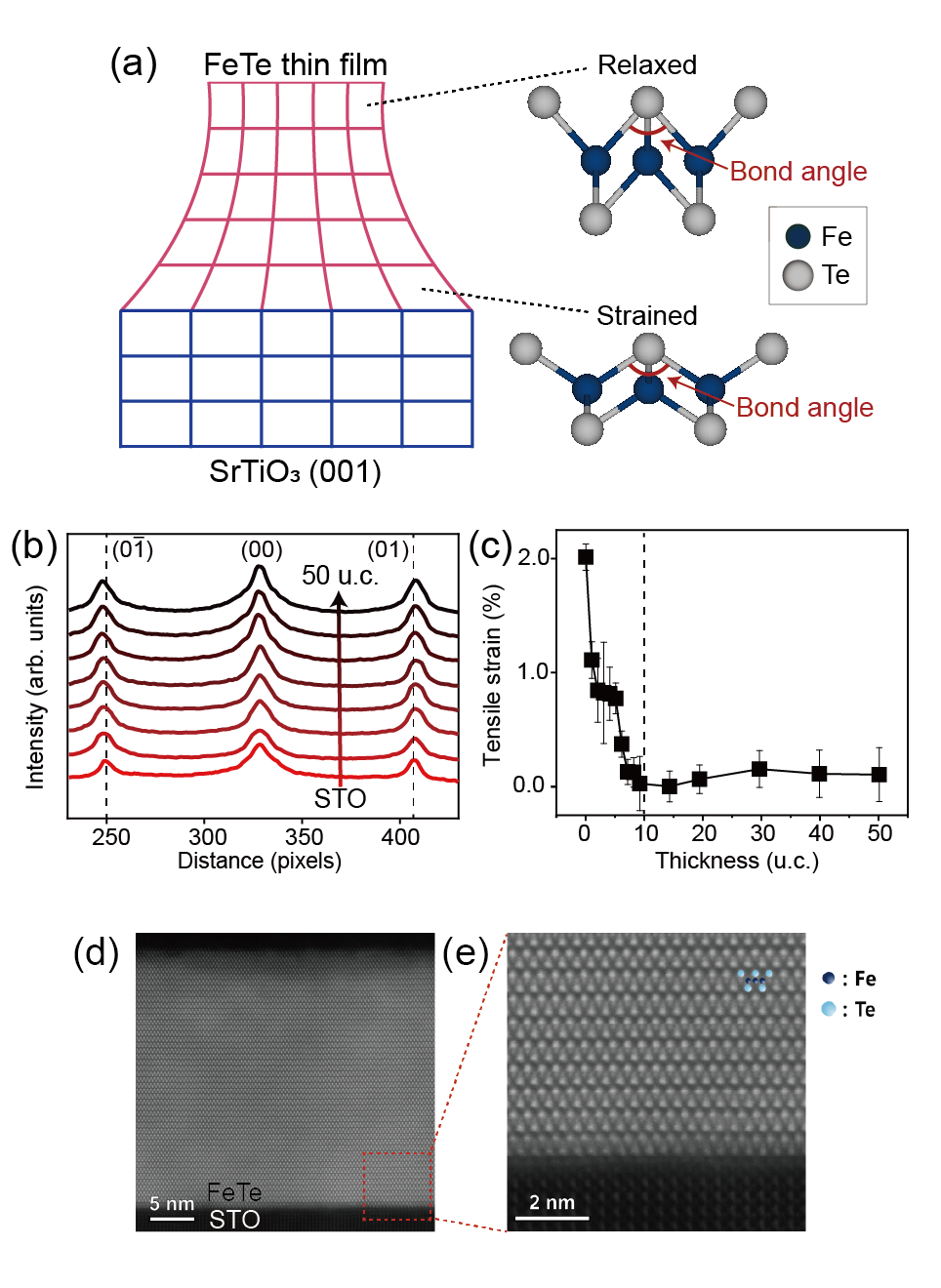}
\vspace{-0.5cm}
\caption{{\bf FeTe thin films under tensile strain.} 
(a) Schematic of strained FeTe thin films on SrTiO$_{\rm 3}$ substrate with local crystal structures. (b) Thickness dependent RHEED pattern profiles measured at 543 K. (c) Magnitude of tensile strain versus film thickness. High-angle annular dark-field (HAADF) STEM image of (d) the overall film and (e) the region near the interface.}
\label{fig:1}
\end{figure}

\begin{figure*}
\includegraphics[width=1\textwidth]{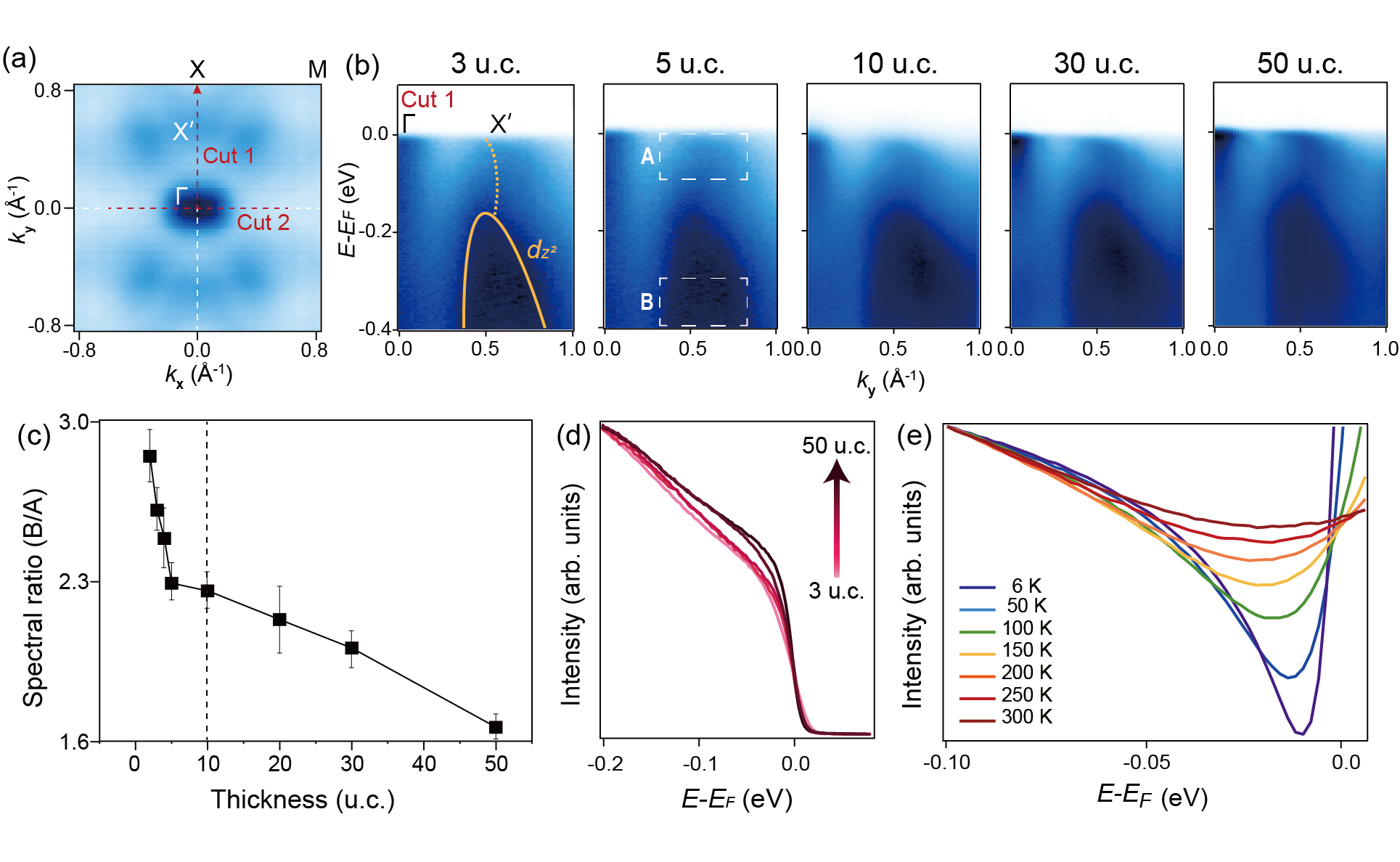}
\vspace{-0.5cm}
\caption{{\bf Thickness dependent $d_{\rm z^{2}}$ evolution in FeTe at the $X'$ point.} 
(a) Fermi surface map of 50-unit-cell FeTe film measured at 6 K. (b) $\Gamma$-$X$ high-symmetry cuts along $k_{\rm x}$ = 0 (Cut 1 in (a)) for various thicknesses. (c) Spectral weight ratio between regions A and B as a function of the film thickness. (d) Energy distribution curves (EDCs) obtained by momentum integration in the range of 0.4~$\mathring{A}^{-1} < k_{\rm y} <$ 0.8~$\mathring{A}^{-1}$. They are normalized to the value at  -0.2 $eV$. (e) Temperature dependent EDCs of 3-unit-cell FeTe film at $X'$ divided by the Fermi-Dirac function, normalized to the value at -0.1 $eV$.}
\label{fig:2}
\end{figure*} 

\section{Results and discussion}
FeTe is a parent compound of iron-based superconductors with a layered tetragonal crystal structure as depicted in Fig. 1(a). The electronic structure of FeTe is mostly determined by three Fe $t_{\rm 2g}$ orbitals with their energies being sensitively dependent on the Fe-Te-Fe bond angle~\cite{yi2017role}. In bulk FeTe, the $d_{\rm xy}$ orbital is expected to be the most strongly renormalized among the three orbitals, leading the system toward an OSMP~\cite{yi2015observation,ye2014extraordinary,PhysRevB.86.195141}. Indeed, previous ARPES studies demonstrated that the $d_{\rm xy}$ orbital becomes completely localized, leaving only incoherent state near the Fermi level~\cite{yin2011kinetic,huang2022correlation}. When FeTe film is grown on a substrate with a mismatched lattice constant, it typically starts to conform to the lattice of the substrate which exerts epitaxial strain. However, as the film thickens, it eventually relaxes to its bulk lattice constant as schematically illustrated in Fig. 1(a). We note that the structural transition is suppressed when FeTe is in the form of a thin film~\cite{sato2025superconductivity, PhysRevLett.104.017003} (See Fig. S1 in the Supplementary Material~\cite{SM} for details.) The strain state of the topmost layer in FeTe film grown on SrTiO$_{\rm 3}$ (STO) substrate can be directly visualized by examining the reflection high energy electron diffraction (RHEED) patterns measured during the growth. Figure 1(b) depicts the extracted profiles of RHEED patterns for varying thicknesses. The distance between the two Bragg peaks (0$\bar{1}$) and (01) increases with the thickness, indicating the decreasing in-plane lattice constant due to the relaxation of the tensile strain.


\begin{figure*}
\includegraphics[width=1\textwidth]{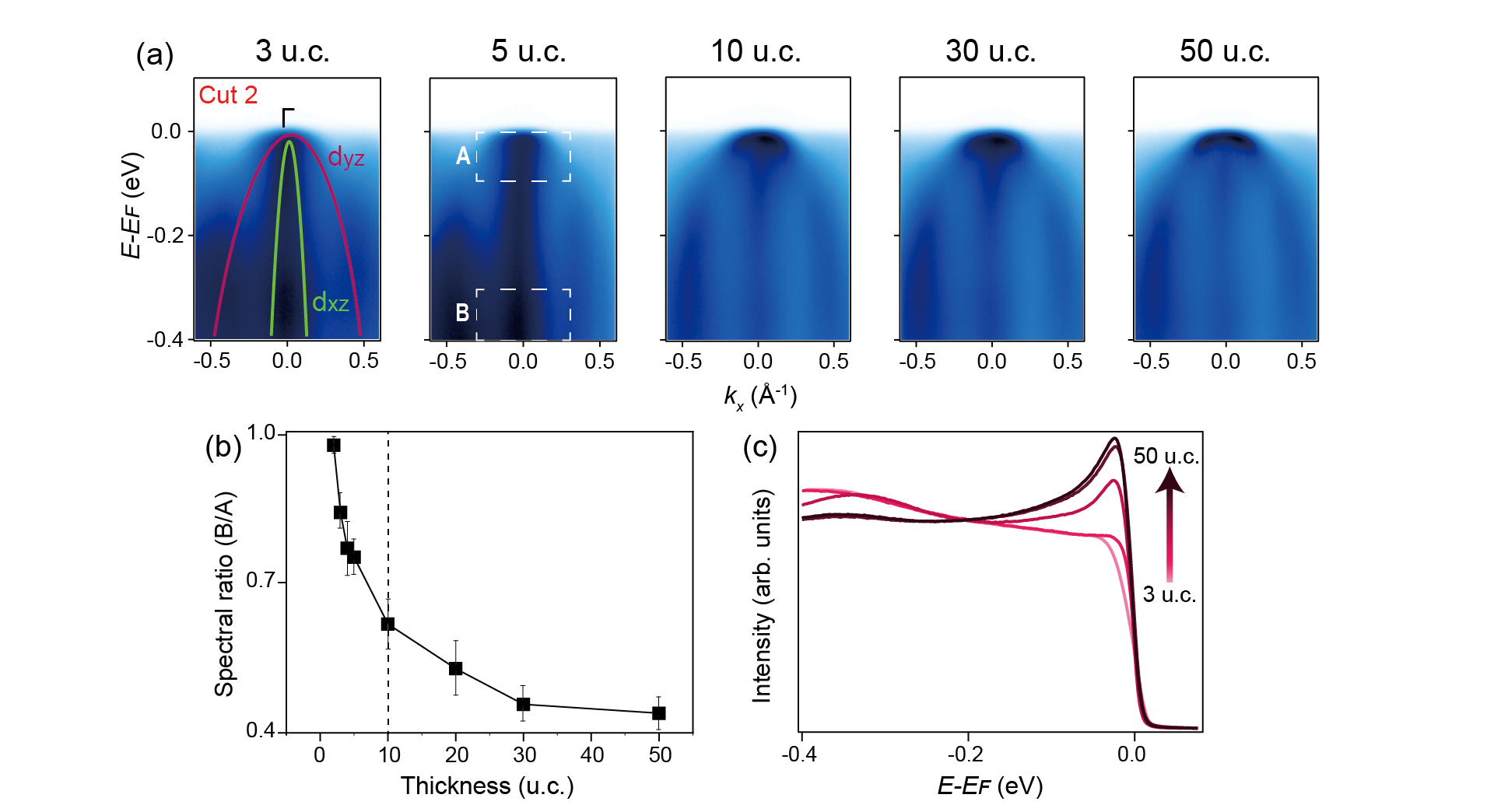}
\vspace{-0.5cm}
\caption{{\bf Thickness dependent $d_{\rm xz}$ band at the $\Gamma$ point.} 
(a) $\Gamma$-$X$ high-symmetry cuts along $k_{\rm y}$ = 0 (Cut 2 in Fig. 2(a)). (b) Spectral weight ratio between from A and B regions as a function of the film thickness. (c) EDCs obtained by momentum integration in the range of -0.2~$\mathring{A}^{-1} < k_{\rm x} <$0.2~~$\mathring{A}^{-1}$, normalized to the value at -0.2 $eV$.}
\label{fig:3}
\end{figure*} 
For quantitative analysis of the strain relaxation, we plot the degree of tensile strain computed from the diffraction peak spacing as a function of thickness in Fig. 1(c). We observe a rapid strain relaxation - few initial layers are subject to significant tensile strain up to 2 \%, but the strain fully relaxes beyond 10 layers. Cross-sectional scanning transmission electron microscopy (STEM) measurements in Fig. 1(d) and 1(e) confirm high crystalline quality across the film as well as an abrupt interfacial region. The atomic peak distance extracted from STEM image clearly reveals the progression of strain relaxation throughout the film (Fig. S2~\cite{SM}), which could not be resolved by X-ray reciprocal space mapping (Fig. S3~\cite{SM}). This rapid strain relaxation rate of FeTe thin film is consistent with or even faster than what has been reported for FeSe thin film, where the strain persists up to 35-unit-cell~\cite{tan2013interface}. The weak inter-layer coupling between the FeTe layers allows the film to relax quickly from the strain imposed by the STO substrate~\cite{ricci2013van}. Identifying the maximum thickness over which strain remains effective is crucial for our electronic structure studies.

The correlation between the thickness of the FeTe film and the effective strain imposed on the surface layer in FeTe thin films allows us to conduct ARPES measurements to study the systematic evolution of their electronic structure with strain. Being a surface-sensitive probe, ARPES is ideally suited to investigate strain effects tuned by the film thickness. The Fermi surface map in Fig. 2(a) shows no dramatic change with varying thickness. For all thicknesses, a strong, well-defined hole pocket is present at the $\Gamma$ point. Weaker spectral weight is distributed near the $X$ and $M$ points, a characteristic feature of FeTe~\cite{kim2023kondo,lin2013nature,PhysRevLett.103.037002,liu2013measurement} (Fig. S4~\cite{SM}). While the overall fermiology remains consistent, significant differences in the band dispersions are observed depending on the thickness. Figure 2(b) presents thickness-dependent high-symmetry cuts ($\Gamma$-$X$) along $k_{\rm x}$ = 0 (‘Cut 1’ in Fig. 2(a)). The spectral weight near the $X$ point (denoted as $X'$), which can be attributed to the $d_{\rm z^{2}}$ band, is redistributed toward the Fermi level as the thickness increases (and tensile strain relaxes). Here, the spectral weight transfer serves as a proxy for the band shift. In the 3-unit-cell film, the $d_{\rm z^{2}}$ band near the Fermi level is weak - as indicated by the dotted line in Fig. 2(b) - yet it becomes increasingly pronounced in thicker films. A similar tendency is reported in Se-doped FeTe that the $d_{\rm z^{2}}$ band is redistributed toward the Fermi level as the $d_{\rm xy}$ orbital becomes localized due to the de-hybridization between two orbitals~\cite{huang2022correlation,yu2017orbital}. Thus, the emergence of the $d_{\rm z^{2}}$ band near the Fermi level is a proxy of the orbital-selective Mott phase of the $d_{\rm xy}$ band. This provides compelling evidence for the breakdown of the OSMP in tensile-strained FeTe, in contrast to the strongly localized d orbital in relaxed FeTe, which exhibits an OSMP with a complete loss of spectral weight near the Fermi level.

To quantify the breakdown of the OSMP, we evaluated the spectral weight ratio between two energy regions. By plotting the ratio of spectral weight near the Fermi level (region A, from 0 to -100 $meV$) and a higher binding energy region (region B, from -300 to -400 $meV$), a notable deceleration is observed at approximately a 10-unit-cell thickness (Fig. 2(c)). This finding is in good agreement with the behavior of the RHEED patterns in Fig. 1(c). A monotonic decrease in the spectral weight ratio persists even above 10-unit-cell, which can be attributed to the gradual improvement of crystallinity with increasing thickness. Nonetheless, Fig. 1(e) demonstrates that the films in the thin-film limit are free of defects and any dislocations. To further capture this evolution, we analyzed the spectral weight transfer of the momentum-integrated energy distribution curves (EDCs) at $X'$ for different thicknesses. As Fig. 2(d) shows, the $d_{\rm z^{2}}$ band with a slight gap progressively fills in as the thickness increases. This is a direct consequence of the OSMP, where the previously gapped-out orbital now contributes to the electronic states at the Fermi level~\cite{huang2022correlation}. 

Temperature-dependent measurements also show further evidence for the OSMP. It is well established that raising temperature drives the iron chalcogenides to an OSMP~\cite{wang2014orbital,ding2014strong,li2014mott,pu2016temperature}. To investigate whether this effect persists in tensile-strained FeTe, the temperature-dependent EDC of a strained 3-unit-cell film was measured at $X'$ (Fig. 2(e)). The EDC is divided by the Fermi-Dirac distribution convolved with a Gaussian to eliminate the effects of thermal broadening. The spectral weight of the integrated EDCs at the $X'$ point is also redistributed closer to the Fermi level as temperature decreases. The evolution of the $d_{\rm z^{2}}$ band indicates that OSMP in bulk-like FeTe collapses not only due to tensile strain but also with increasing temperature, similarly to Se-doped FeTe~\cite{yi2013observation,huang2022correlation}.

While strained FeTe and Se-doped FeTe exhibit similar trends in the $d_{\rm xy}$ band~\cite{huang2022correlation}, they display opposite behaviors in the $d_{\rm xz}$ bands. Figure 3(a) presents thickness-dependent high-symmetry cuts ($\Gamma$-$X$) along the $k_{\rm y}$ = 0 ('Cut 2' in Fig. 2(a)). At the $\Gamma$ point, two hole-like bands appear - the inner/outer band corresponds to $d_{\rm xz}$/$d_{\rm yz}$ orbital character~\cite{morfoot2023resurgence,PhysRevB.81.014526}. In strained films, the spectral weight of $d_{\rm xz}$ is distributed over a wide range of binding energies. As tensile strain relaxes, the overall spectral weight is redistributed towards the Fermi level. We plotted the ratio of spectral weight between regions A and B at $\Gamma$ point in Fig. 3(b). We found a logarithmic drop in the spectral ratio as the thickness is increased, which also aligns well with the degree of tensile strain measured from RHEED. While Fig. 3(b) does not display a pronounced kink around 10-unit-cell as Fig. 2(c) does, the effective mass of the $d_{\rm xz}$ orbital exhibits a distinct crossover at a similar thickness (Fig. S5~\cite{SM}). The consistency between the effective mass evolution and the strain relaxation behavior strongly supports a close link between the effect of tensile strain and the electronic evolution of the $d_{\rm xz}$ orbital. Spectral transfer from higher binding energies to Fermi level is well resolved in Fig. 3(c), which shows EDCs at $\Gamma$. In relaxed thick films, a sharp quasi-particle peak near the Fermi level is observed as a manifestation of the $d_{\rm xz}$. 

\begin{figure}
\includegraphics[width=0.48\textwidth]{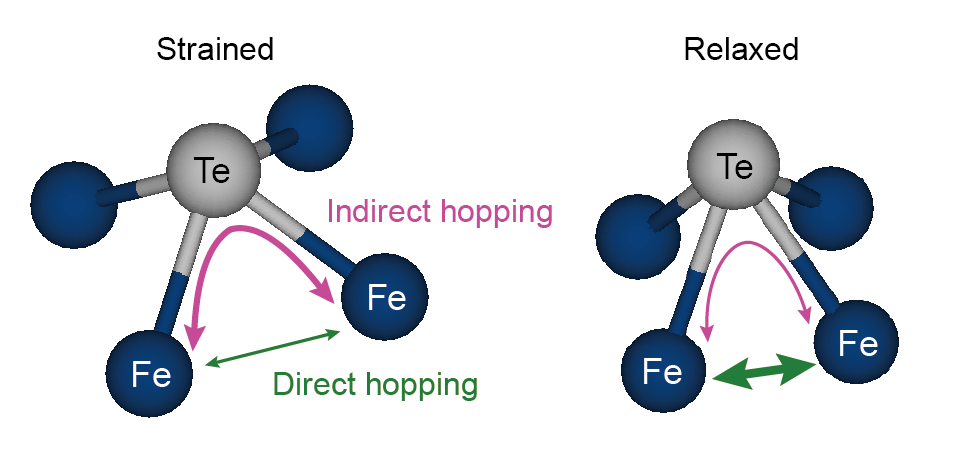}
\vspace{-0.5cm}
\caption{{\bf Hopping mechanisms upon tensile strain.} 
(Left) Schematic of tensile-strained FeTe thin films with suppressed direct hopping. (Right) Schematic of relaxed FeTe thin films with enhanced direct hopping. Indirect hopping remains nearly unchanged.}
\label{fig:1}
\end{figure}

The suppression of the well-defined quasi-particle peak in the $d_{\rm xz}$ band under tensile strain in Fig. 3(c) are unexpected~\cite{huang2022correlation,ieki2014evolution,PhysRevB.96.035137}. In iron-based superconductors, it is well established that correlation strength increases as the Fe–Ch–Fe bond angle decreases~\cite{yi2017role}. The observed enhancement of correlations in the $d_{\rm xz}$ orbitals under tensile strain therefore suggests that an additional control parameter is at play, since the Fe–Ch bond length is essentially fixed in tensile-strained FeTe. As for the underlying cause, dimensionality and defect can be ruled out based on the evidence provided in Fig. S5~\cite{SM} and Fig. 1(e).

One possible way to explain this is to trace how the structural parameter modifies the hopping pathways. Tensile strain enlarges the in-plane Fe–Fe lattice constant, which suppresses the overlap between the orbitals and therefore weakens the direct Fe–Fe hopping for the $d_{\rm xz}$ orbitals~\cite{yin2011kinetic} as illustrated in Fig. 4. This reduction in hopping integral effectively enhances the correlation, leading to reduced coherence even though a larger Fe–Ch–Fe bond angle would normally enhance coherence. Because the Fe–Ch bond length remains essentially unchanged, the indirect hopping channel is largely unaffected, leaving the suppression of direct Fe–Fe hopping as the dominant effect. In the case of Se doping, in contrast, changes in the Fe–Fe spacing are inseparable from variations in the Fe–Ch bond length, which simultaneously modifies the indirect channel~\cite{yi2017role}. Such entanglement has likely obscured the distinct role of the Fe–Fe lattice constant. Epitaxial strain, by partially decoupling these parameters, makes clear that the in-plane Fe–Fe distance is a key structural knob governing the coherence of the $d_{\rm xz}$ states.

\section{Conclusions and outlook}
We have successfully employed tensile strain to manipulate the physical properties of FeTe thin films. Using {\it in-situ} RHEED and ARPES, we established a qualitative correlation between strain relaxation and the evolution of the electronic band structure. RHEED measurements demonstrate that the strain undergoes a logarithmic relaxation, becoming fully relaxed after approximately 10-unit-cell. ARPES captures the electronic structure at different stages of strain relaxation. We provide clear evidence for the OSMP through the development of the $d_{\rm z^{2}}$ orbital at the $X'$ point. Our data reveal that this behavior is not only thickness-dependent but also temperature-dependent, suggesting that the OSMP is an intrinsic feature of FeTe. Moreover, we observed that the spectral weight of the $d_{\rm xz}$ orbital at the $\Gamma$ point approaches the Fermi level with increasing thickness. Of particular interest are the emergent quasi-particle peak and coherence of the $d_{xz}$ band, in contrast to those reported in Se-doped bulk FeTe~\cite{huang2022correlation,ieki2014evolution,PhysRevB.96.035137}. By examining the hopping mechanisms, we show that tuning the bond angle via strain produces outcomes distinct from chemical doping, direct Fe-Fe hopping being a contributing factor controlling the orbital-dependent electronic correlations.

Superconductivity in tensile-strained FeTe thin films and FeTe-based heterostructures has been widely reported in recent years, suggesting that strain and interfacial effects can significantly control the intricate interplay between magnetic order and superconductivity~\cite{sato2025higher,he2014two,manna2017interfacial,tkavc2024multiphase,yi2025universal,ciechan2013magnetic,song2024growth,PhysRevLett.104.017003}. Bulk FeTe was knwon as a non-superconducting antiferromagnet ~\cite{johnston2010puzzle,glasbrenner2015effect,ma2009first}, but a recent study reverses this long-standing consensus and establishes that stoichiometric FeTe is a superconductor with the suppressed long-range magnetism~\cite{yan2026stoichiometric}. Given that the magnetic order and pairing symmetry in iron-based chalcogenides are intimately coupled to lattice and orbital degrees of freedom, tuning the lattice structure becomes a direct pathway to control these competing states. In this regard, our strain-dependent study demonstrates how lattice deformation modulates the orbital-dependent correlations, thereby providing a microscopic mechanism for the suppression of magnetism and the subsequent emergence of superconductivity in FeTe-based systems.

\section{Methods}
Epitaxial FeTe thin films were grown on (001)-oriented Nb-doped STO (0.05 wt\%) substrates via molecular beam epitaxy (MBE) system. Based on the lattice constants of FeTe and STO, a tensile strain of 2~\% is estimated for the FeTe thin films. Prior to the growth, 10-unit-cell STO buffer layers were deposited using a pulsed laser deposition (PLD) system on STO substrate as shown in Fig. S6~\cite{SM}. Fe (99.999~\%) and Te (99.999~\%) were codeposited in UHV at growth temperature of 270 $^{\circ}$C with Te flux 7 times greater than that of Fe. Sample quality were characterized using X-ray diffraction and atomic force microscopy (AFM) in Fig. S7~\cite{SM}. As detailed in Fig. S8~\cite{SM}, the thickness variation is significantly reduced in the thin regime, remaining within approximately 1 to 2 unit cells, thereby validating thickness as a well-defined and reliable control parameter.

The STEM lamella sample was prepared by focused ion beam milling using an FEI Helios G4 system, with rough milling performed at 30kV followed by final thinning at 5 kV to minimize surface damage. HAADF STEM imaging was performed using a Thermo Fisher Scientific Themis Z aberration-corrected and monochromated miscroscope, with the accelerating voltage set up to 200 kV. The probe current was set to 80 pA with a dwell time of 2 $\mu$s, and HAADF images were collected over a semi-angle range of 54-200 mrad.

{\it In-situ} ARPES measurements were performed at 6 K using the home-built laboratory system equipped with a Scienta DA30 analyzer and a discharge lamp from the Fermi instrument. He-$I\alpha$ ($hv$ = 21.2 eV) light partially polarized in a linear vertical direction was used.

\vspace{0.5cm}
\acknowledgments
We thank Dongjin Oh, Youngdo Kim, and Byungmin Sohn for enlightening discussions. This work was supported by the National Research Foundation of Korea (NRF) grant funded by the Korea government (MSIT). (No. 2022R1A3B1077234). This research was also supported by Basic Science Research Program through the National Research Foundation of Korea (NRF) funded by the Ministry of Education (No. RS-2019-NR040081) and the Institute of Applied Physics of Seoul National University.

%

\end{document}